\documentclass[journal]{IEEEtran} 
\IEEEoverridecommandlockouts

\usepackage{cite}
\usepackage{amsmath,amssymb,amsfonts}
\usepackage{algorithmic}
\usepackage{graphicx}
\usepackage{textcomp}
\usepackage{xcolor}
\usepackage{placeins}

\usepackage{url}  

\def\BibTeX{{\rm B\kern-.05em{\sc i\kern-.025em b}\kern-.08em
    T\kern-.1667em\lower.7ex\hbox{E}\kern-.125emX}}
\begin{document}

\title{Dynamics of Resource Allocation in O-RANs: An In-depth Exploration of On-Policy and Off-Policy Deep Reinforcement Learning for Real-Time Applications}

\author{
    \IEEEauthorblockN{Manal Mehdaoui$^1$, }
    \IEEEauthorblockN{Amine Abouaomar$^1$ }
    \\
    \IEEEauthorblockA{\textit{$^1$ School of Science and Engineering, Al Akhawayn University in Ifrane, Morocco.}}\\
    
}

\maketitle

\begin{abstract}

Deep Reinforcement Learning (DRL) is a powerful tool used for addressing complex challenges in mobile networks. This paper investigates the application of two DRL models, on-policy and off-policy, in the field of resource allocation for Open Radio Access Networks (O-RAN). The on-policy model is the Proximal Policy Optimization (PPO), and the off-policy model is the Sample Efficient Actor-Critic with Experience Replay (ACER), which focuses on resolving the challenges of resource allocation associated with a Quality of Service (QoS) application that has strict requirements.

Motivated by the original work of Nessrine Hammami and Kim Khoa Nguyen, this study is a replication to validate and prove the findings. Both PPO and ACER are used within the same experimental setup to assess their performance in a scenario of latency-sensitive and latency-tolerant users and compare them. The aim is to verify the efficacy of on-policy and off-policy DRL models in the context of O-RAN resource allocation.

Results from this replication contribute to the ongoing scientific research and offer insights into the reproducibility and generalizability of the original research. This analysis reaffirms that both on-policy and off-policy DRL models have better performance than greedy algorithms in O-RAN settings. In addition, it confirms the original observations that the on-policy model (PPO) gives a favorable balance between energy consumption and user latency, while the off-policy model (ACER) shows a faster convergence. These findings give good insights to optimize resource allocation strategies in O-RANs.

\end{abstract}


\section{Introduction}
Reinforcement Learning (RL) is an approach that examines the dynamic interaction between agents and environments, which aims to achieve the highest possible cumulative reward. The mix of RL with Deep Neural Networks (DNN) has led to the technique of Deep Reinforcement Learning (DRL) in addressing high-complexity problems. DRL includes two model types: on-policy and off-policy and they have different characteristics and advantages [1]. On-policy models like Proximal Policy Optimization (PPO), are known for stability and ease of implementation but suffer from data inefficiency. In contrast, off-policy models, for instance, Sample Efficient Actor-Critic with Experience Replay (ACER), show data efficiency but often face stability challenges [2].
The usage of both on-policy and off-policy DRL models in solving resource allocation problems within mobile networks has given a critical examination of their performance. While past works have applied singular DRL techniques for resource allocation in scenarios like cloud RANs and multi-tenant networks, a comparative analysis of on-policy and off-policy models in the context of real-time surveillance video resource allocation remains important [3]. 
Motivated by this research, our study is a replication study of ''On-Policy vs. Off-Policy Deep Reinforcement Learning for Resource Allocation in Open Radio Access Networks.''. Our goal is to investigate how effective the on-policy (PPO) and off-policy (ACER) DRL models are in the scenario of dynamic slicing within Open Radio Access Networks (O-RAN). It focuses on real-time surveillance video with the metric of latency as important for Quality of Service (QoS). The use of wireless video surveillance cameras (VSCs) in domains of public safety and intelligent transportation systems requires innovative solutions to meet the QoS requirements[4]. Many researchers addressed resource allocation in video surveillance, however, they mainly focused on a single type of DRL model without comparing the effectiveness of on-policy and off-policy approaches [5]. In reference [6], the authors examined video surveillance within construction sites, where computational and networking resources are frequently constrained. The proposal was an edge-based approach using a graph-assisted hierarchical Deep Q-Network (DQN) algorithm to ensure both dependable accuracy and minimal delay.
The goal of this paper is to answer: Which type of DRL models, on-policy or off-policy, is more efficient in solving the resource allocation problem for real-time surveillance video in O-RANs? To answer this, we use the O-RAN architecture. This architecture is known for its openness, intelligence, and interoperability. The O-RAN architecture provides a suitable platform for hosting AI workflows, especially DRL models. Our contributions include the evaluation of PPO and ACER in finding an efficient network resource allocation strategy for real-time video surveillance in a dynamic slicing O-RAN system, considering both latency-sensitive and latency-tolerant users. We formulate the problem as a Mixed-Integer Programming (MIP) model, providing optimal bounds to assess the performance of on-policy and off-policy DRL models.
The remainder of this paper is organized as follows: Section II provides a review of related work to gain more understanding and explore different approaches to our problem. Section III, which is the System Model talks about the O-RAN architecture and problem formulation. Section IV introduces the DRL-based resource allocation solution or the proposed solution, followed by the Evaluation and results in Section V, where we evaluate the models' performance and results. Finally, we conclude our findings in Section VI.
\section{Related Work}

In the topic of Dynamics of Resource allocation in O-RANs, many studies have explored and covered many techniques and approaches to find a solution to different problems and challenges found in this domain. In this paper [7] authors propose a UK-means-based clustering and deep reinforcement learning (DRL) approach to address localization uncertainties and dynamic traffic demands in 5G mmWave networks. Moreover, this study [8] examines how mixing on-policy Monte Carlo updates with off-policy, bootstrap Q-Learning updates affects the performance of deep reinforcement learning methods. The results point out partial improvements in some scenarios and give insight into reinforcement learning optimization strategies. Additionally, a study [9] investigates reinforcement learning optimization strategies, which contributes to the ongoing exploration of effective resource allocation techniques in O-RANs. Furthermore, an additional study [10] explores mobility-aware load balancing for optimizing O-RAN networks using machine learning algorithms. The proposed multi-agent multi-armed bandit for load balancing and resource allocation (mmLBRA) scheme aims to distribute loads across the network, to prevent congestion and improve the effective sum-rate performance of the O-RAN network. Simulation results show significant enhancements in network sum rate and load balancing compared to existing heuristic methods. These studies give insight into different methods of optimizing resource allocation in O-RANs, tackling challenges around QoS, network performance, and congestion prevention.

\section{System Model}
A. System model

In Fig1, we use a sliced O-RAN system for our replication study. This system is composed of N O-RUs linked to a pool of M virtualized O-DUs (vO-DUs) These vO-DUs, implemented as virtual network functions (VNFs), operate in a cloud environment. End-users with varying Quality of Service (QoS) needs can connect to a nearby O-RU, gaining access to network services through wireless communication links.

\begin{figure}[htbp]
    \centering
    \includegraphics[width=0.9\linewidth]{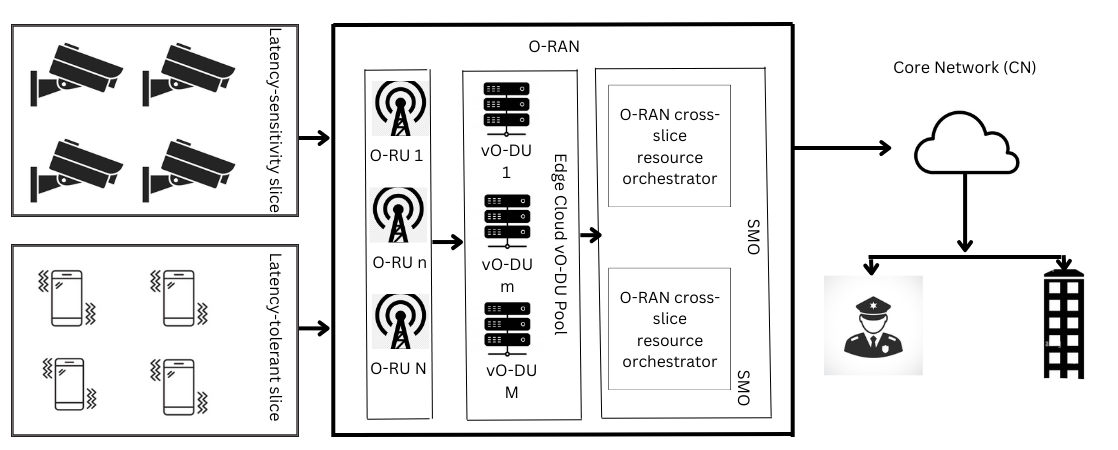}
                                                {   Fig.1 System model} 
       
\end{figure}

We are considering two service types, which are distinguished by their latency characteristics. The first is a latency-sensitive service with a specified maximum latency constraint, and the second is a latency-tolerant service. Accordingly, two slices are established: one catering to VSCs for real-time video transmission via O-RAN to the 5G Core Network (CN) and the other to a Control Center (CC) like a police station for real-time monitoring; the other serving latency-tolerant users. 

These slices are overseen by an O-RAN cross-slice resource orchestrator hosted by the SMO. The set of VSCs is represented as W, and the set of latency-tolerant users is represented as V. The demands of VSC w and latency-tolerant user v in computing resources are marked as cw and bv, respectively, for all $w \in W$ and $v \in V$. Computing resource capacity, measured by the number of CPUs, assumes a fixed frequency F for all CPUs.

The system operates in discrete time with T-slicing windows. At the start of each slicing window, the O-RAN cross-slice resource orchestrator determines and executes resource allocation actions, and updates the O-RAN state. Throughout the remainder of the slicing window, traffic remains stationary, and resource allocation decisions remain unchanged. A resource allocation decision dictates the association of an end-user with a vO-DU.

So we have two following decision variables defined:
 
\[
\begin{aligned}
(1) \quad x^t_{m,w} &=
\begin{cases} 
1, & \text{if } c_w \text{ of vO-DU } m \\
& \text{ is allocated to VSC } w \\
& \text{ during slicing window } t, \\
0, & \text{otherwise.}
\end{cases}
\end{aligned}
\]

\[
\begin{aligned}
(2) \quad y^t_{m,v} &=
\begin{cases} 
1, & \text{if } b_v \text{ of vO-DU } m \\
& \text{ is allocated to user } v \\
& \text{ during slicing window } t, \\
0, & \text{otherwise.}
\end{cases}
\end{aligned}
\]
\

The CPU usage of a vO-DU at a given slicing window is computed as the sum of the computing resources allocated to VSCs and latency-tolerant users.

\[
\begin{aligned}
(3) \quad z^t_{,m} &= \sum_{w=1}^{W} x^t_{m,w} c_w + \sum_{v=1}^{V} y^t_{m,v} b_v
\end{aligned}
\]

We assume that the traffic from latency-tolerant users follows an M/M/1 queue with an arrival rate $\lambda_v$ for each user $v$. This results in a total arrival rate of latency-tolerant users at a vO-DU equal to the sum of the arrival rates for all users in the set $V$.

\[
(3.1) \quad \sum_{v=1}^{V} \lambda_v
\]

To make it simple, we assume that each Video Surveillance Camera (VSC) monitors a specific area continuously and stores data in a First-In-First-Out (FIFO) queue. All VSCs are the same and capture videos of fixed size and duration at a fixed rate $\lambda$, which are then processed at a fixed service rate $\mu$. We consider $\mu$ to be greater than $\lambda$ to ensure negligible queueing delay in the VSC queue. The transmission of VSC packets from an O-RU to a vO-DU can be modeled as a deterministic D/D/1 queue with an arrival rate $\mu$. The total arrival rate of VSC packets at a vO-DU equals the sum of arrival rates for all VSCs in the set $W$.

The latency of a VSC packet includes the propagation delay, queueing delay, and computing delay.
\[
(4) \quad D_w = DP_{{m,n}} + DQ_{{m,w}} + DC_{{m,w}}
\]

The propagation delay from an O-RU to a vO-DU is determined by the distance between them divided by the signal propagation speed.

\[
(5) \quad DP{{m,n}} = \frac{d_{n,m}}{s}
\]

The queueing delay experienced by a VSC packet is defined in terms of the vO-DU service rate and the total packet arrival rate at the vO-DU.

\[
(6) \quad DQ_{{m,w}} = \frac{1}{(z^t_{m}F) - (W \mu + \sum_{v=1}^{V} \lambda_v)}
\]

The computing delay for processing a VSC packet in a vO-DU depends on the packet size and the vO-DU service rate.
\[
(7) \quad DC_{{m,w}} = \frac{\beta_w}{z^t_{m}F}
\]

B. Problem Formulation

The evaluation of the performance of the system model is based on the definition of the power consumption of the vO-DU pool. This is calculated as follows at a specific time slice ''t'' for an active vO-DU as ''m'' [11]:

\[
P_t^m = P(0\%) + \left(P(100\%) - P(0\%)\right)\left(2z_t^m - (z_t^m)^{1.4}\right)
\]

Here, \(P(0\%)\) and \(P(100\%)\) represent the power consumption of vO-DU ''m'' in idle mode and under full load, respectively. All \(M\) vO-DUs share identical values for \(P(0\%)\) and \(P(100\%)\).

The objective is to assign end-users to vO-DUs in a manner that minimizes the vO-DU pool's power consumption while adhering to Quality of Service (QoS) requirements for various types of end-users. Therefore, the problem is framed as:

\[
\text{minimize} \sum_{t=1}^{T} \sum_{m=1}^{M} P_t^m
\]

Subject to the following constraints:

\[
z_t^m \leq Z_{\text{max}}^m, \forall m \in M \quad (9a)
\]

\[
D_w < D_{\text{th}}, \forall w \in W \quad (9b)
\]

\[
(z_t^m F) - \left(W \mu + \sum_{v=1}^{V} \lambda_v\right) \geq 0, \forall m \in M \quad (9c)
\]

Here, \(Z_{\text{max}}^m\) is the maximum computing capacity of vO-DU ''m.'' Constraint (9a) ensures system stability by limiting the total allocated computing resources in each vO-DU. Constraint (9b) specifies that the latency experienced by each VSC packet must be below the maximum tolerable latency for latency-sensitive slices. Constraint (9c) addresses queueing stability for latency-tolerant users. This allocation problem is formulated as a Mixed Integer Programming (MIP) model.

\section{Proposed Solution}
\section*{DRL-Based Computing Resource Allocation}

The decision process for allocating computer resources has been conceptualized as a Markov Decision Process (MDP). MDP is represented as a five-tuple \(< S, A, P(s_0|s, a), R, \gamma >\), where:

\begin{itemize}
    \item \(S\) denotes a finite state space.
    \item \(A\) represents a finite action space.
    \item \(P(s_0|s, a)\) signifies the transition probability from state \(s \in S\) with action \(a \in A\) to state \(s_0 \in S\).
    \item \(R(s, a)\) is the immediate reward obtained from taking action \(a\) in state \(s\).
    \item \(\gamma \in [0, 1]\) is the discount factor, reflecting the significance of future rewards.
\end{itemize}

In this system, the DRL agent functions as the O-RAN cross-slice resource orchestrator. At each slicing window \(t\), the DRL agent observes the state \(s(t)\) from the state space \(S\), seeks an optimal policy, and selects action \(a(t)\) from action space \(A\) to maximize the system's reward \(R(s, a)\).

\begin{enumerate}[a)]
    \item \textbf{State space}: The state space encompasses the demands of Virtualized Network Functions (VSCs) and latency-tolerant users regarding computing resources, along with network resource utilization. At slicing window \(t\), the observation \(s(t)\) of the DRL agent is defined as:
    
    \[s(t) = [c1, ..., cW, b1, ..., bV, zt1, ..., ztM]\]

    \item \textbf{Action space}: During each slicing window \(t\), the DRL agent determines the vO-DU (virtualized O-RAN Distributed Unit) to associate with the current end-user. Thus, action \(a(t)\) from action space \(A\) is the index of the chosen vO-DU:
    
    \[a(t) = m \in \{1, 2, ..., M\}\]

    \item \textbf{Reward}: In Reinforcement Learning (RL), the aim is to maximize the reward function, while the MIP model's objective function (Equation 9) aims to minimize the vO-DU pool's power consumption. Therefore, under state \(s(t)\), the immediate reward function guiding the DRL agent's action \(a(t)\) towards optimizing the objective function can be defined as the negative weighted sum of the instantaneous power consumption of the vO-DU pool:
    
    \[R(s, a) = -\alpha \sum_{m=1}^{M} P_t^m\]

\end{enumerate}

Here, \(\alpha\) is a positive system parameter defining a weight factor for power consumption. To maximize \(R(s, a)\), the DRL agent minimizes power consumption by selecting suitable end-users for vO-DU associations.

To execute the computing resource allocation process, two DRL models are designed and compared in this scenario: Proximal Policy Optimization (PPO) as an on-policy model and Advantage Actor-Critic (ACER) as an off-policy model.

\begin{figure}
    \centering
    \includegraphics[width=1\linewidth]{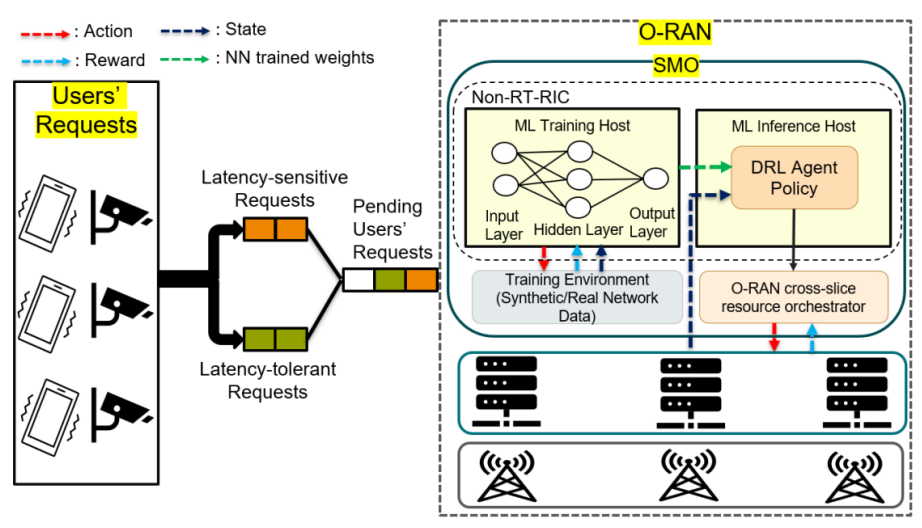}
     {   Fig.2 Resource allocation process} 
    
    \label{fig:enter-label}
\end{figure}

Both models function as model-free RL algorithms since the agent doesn't directly learn the model of the environment or predict future states and values before acting [12]. Instead, the agent approximates the value function for each state-action pair to derive the optimal policy [12]. Both models employ an actor-critic architecture that requires two neural networks (NN), with the state of the environment as input for both.

The actor-network computes a policy function for decision-making, and the critic network evaluates policies, providing feedback on action effectiveness and suggesting adjustments through value function calculations.

PPO [13] is an on-policy model that evaluates policy values while using them for control. It collects a small batch of experiences to update its decision-making policy via policy gradient updates. These experiences are used only once which simplifies implementation complexity. The main advantage of PPO is its ability to address significant policy changes by stabilizing the policy gradient process. And that is by optimizing a clipped surrogate objective function and putting constraints to ensure stable policy updates.

 On the other hand, ACER [14] is an off-policy model that uses a memory replay buffer to store tuples containing state, action, reward, and next-state information. These stored samples are reused for training a value function through multiple gradient-based off-policy updates. The main benefit of ACER is sample efficiency: it learns from all accumulated experiences, and it maximizes the usage of each sampled piece of experience rather than relying on recent ones.

Table I provides a summary of the comparison between the two DRL models.

Figure 2 illustrates the workflow of DRL-based resource allocation in the O-RAN architecture context. When RL is used in O-RAN, it is assumed that ML training and simulation is host co-location, usually, the DRL training process of a Non-Real-Time RAN Intelligent Controller (RIC) takes place in ML within the training host, where the agent of the DRL interacts with the simulated environment. During the training episodes, the DRL operator observes the environment, finds optimal actions, and aims to maximize the reward of the system. The resulting trained model is then put into the ML inference host for dynamic resource allocation in each slicing window \(t\), optimizing incoming end-user requests and updating the environment state accordingly.

\begin{table}[htbp]
    \centering
    \caption{PPO and ACER comparison}
    \label{tab:comparison}
    \begin{tabular}{|l|l|}
    \hline
    \textbf{ACER} & \textbf{PPO} \\ \hline
    Model-free & Model-free \\
    Policy gradient algorithm & Policy gradient algorithm \\
    Off-policy & On-policy \\
    Actor-Critic & Actor-Critic \\
    Sample efficiency & Policy updates stability \\
    Discrete, Continuous environments & Discrete, Continuous environment \\ \hline
    \end{tabular}
\end{table}

\begin{table}[htbp]
    \centering
    \caption{Main Simulation Settings}
    \label{tab:simulation_settings}
    \begin{tabular}{|l|l|}
    \hline
    \textbf{Parameter} & \textbf{Value} \\ \hline
    M & 10 \\ \hline
    \(P(0\%)\) & 87W \\ \hline
    \(P(100\%)\) & 145W \\ \hline
    Actor learning rate & 0.0003 \\ \hline
    Critic learning rate & 0.001 \\ \hline
    \(\gamma\) & 0.99 \\ \hline
    \(\alpha\) & 0.0001 \\ \hline
    \end{tabular}
\end{table}

\section{Evaluation and Results}

In our replication study, we used the available model proposed in the original research and trained them for a duration of three hours. The results are similar to those of the original paper

For the simulation setting, we used the Alibaba cluster-trace-v2018 dataset, which spans an 8-day period and encompasses 4000 machines[15]. The dataset gives information on the server characteristics, latency metrics, and task details. The replication setup is similar to the one of the original paper or paper. This includes the optimal solution derived from the MIP (Mixed integer programming) model which uses the Gurobi solver, and the same for the greedy baseline approach.

The evaluation focuses on different key aspects:
a) Model Performance: We monitored the average episodic reward during the training process, as shown in Figures 3 and 4. There are some variations in the learning rates, but still, both models show an improvement in performance over time. And we observed a faster convergence for the PPO model compared to ACER.

b) Model Stability: Figures 5 and 6 show the training performance of the models under different hyperparameters, particularly the number of neurons in dense layers. While PPO showed consistent performance across different configurations. ACER showed sensitivity to changes, especially with fewer neurons.

c) Energy Consumption: Figure 7 presents a comparison of energy consumption between the models under different levels of network load. Both DRL models showed superior performance compared to the greedy baseline, with PPO showing the most efficient energy consumption, which resembles the optimal solution.

\begin{figure}
    \centering
    \includegraphics[width=1\linewidth]{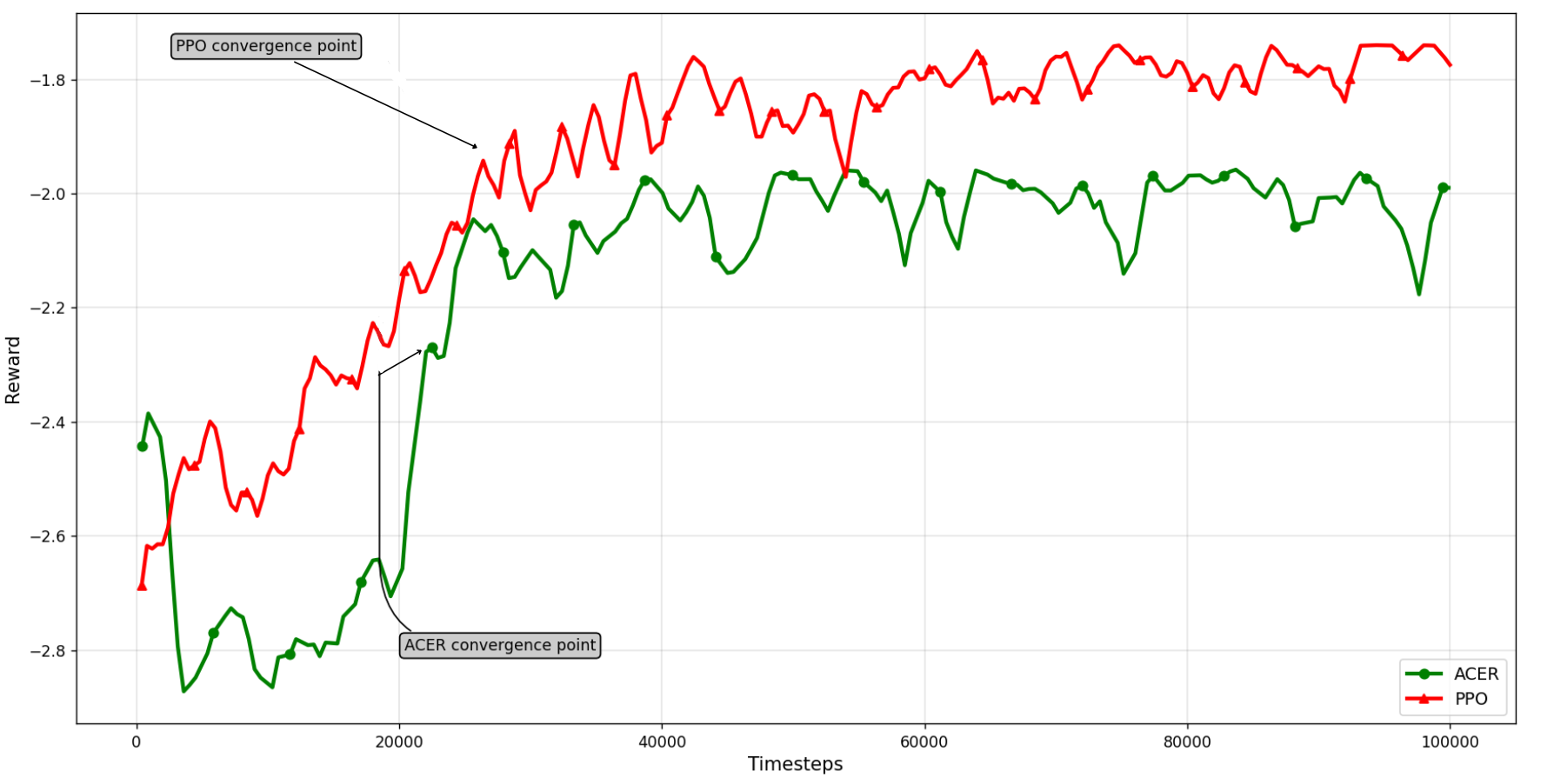}
    {Fig. 3: Steps-reward curve for PPO and ACER} 
    \label{fig:steps-reward}
\end{figure}

\begin{figure}
    \centering
    \includegraphics[width=1\linewidth]{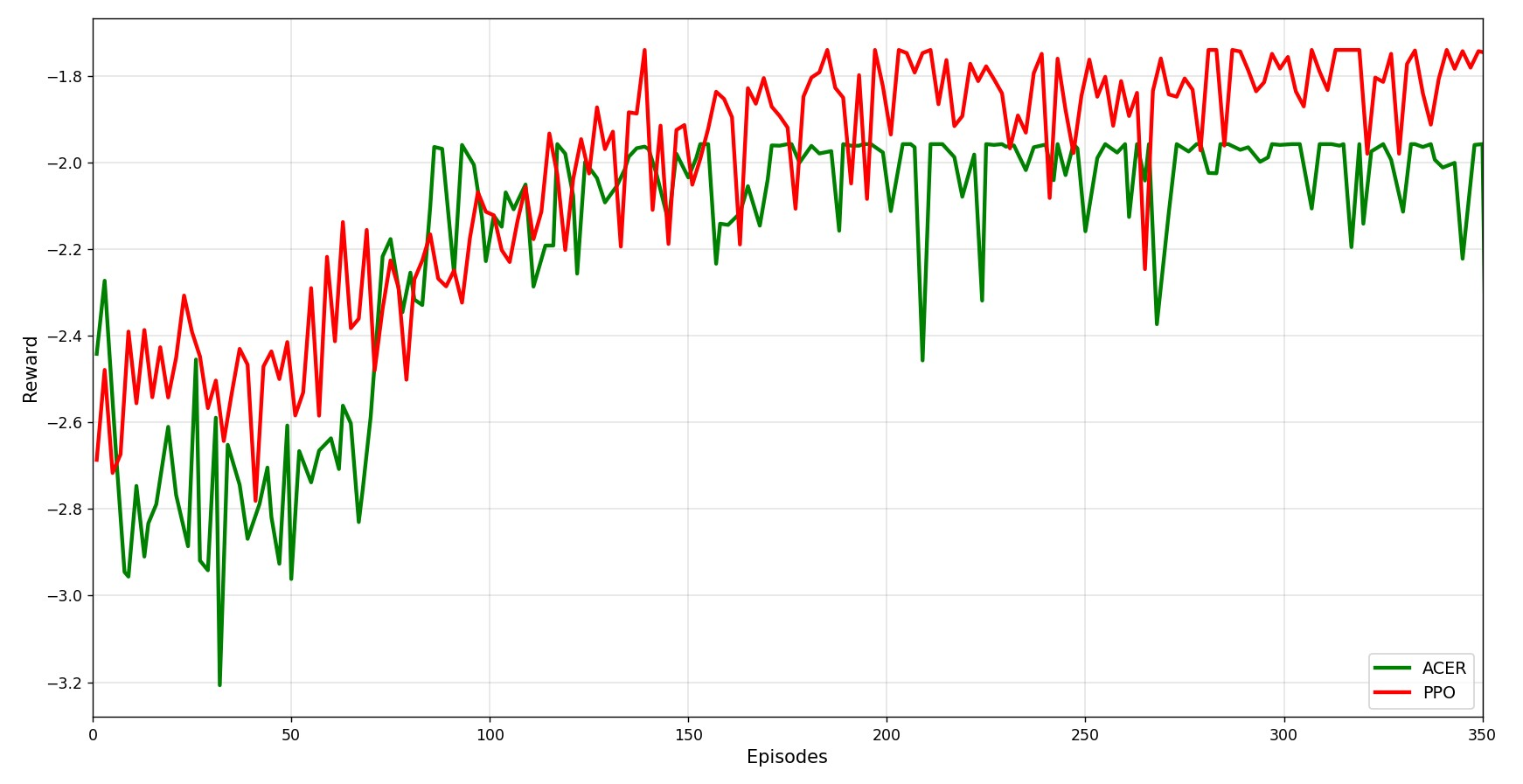}
    {Fig. 4: Episodes-reward curve for PPO and ACER} 
    \label{fig:episodes-reward}
\end{figure}

d) Trade-off Analysis: Figure 8 shows the trade-off between user latency and energy consumption for all models. Both DRL approaches outperformed the greedy baseline. PPO showed better optimization in balancing latency and energy efficiency compared to ACER.

The findings align closely with those of the original study, proving the efficacy of the proposed DRL models in addressing resource allocation challenges in O-RAN systems. Moreover, the replication study shows the robustness and reproducibility of the results across different experimental setups.
\begin{figure}
    \centering
    \includegraphics[width=1\linewidth]{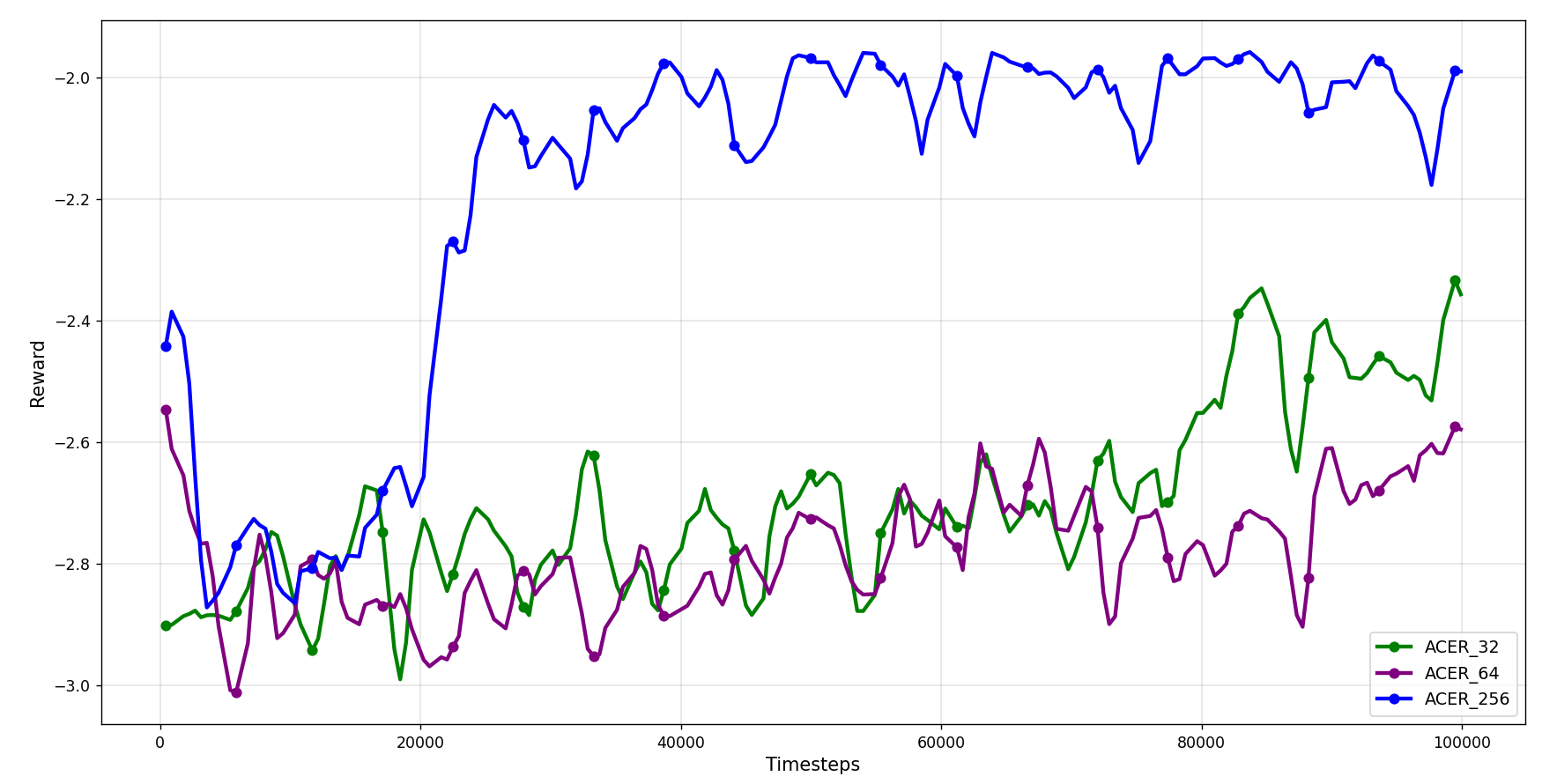}
    {Fig. 5: ACER reward for different NN architectures.}
    \label{fig:acer-arch-reward}
\end{figure}

\begin{figure}
    \centering
    \includegraphics[width=1\linewidth]{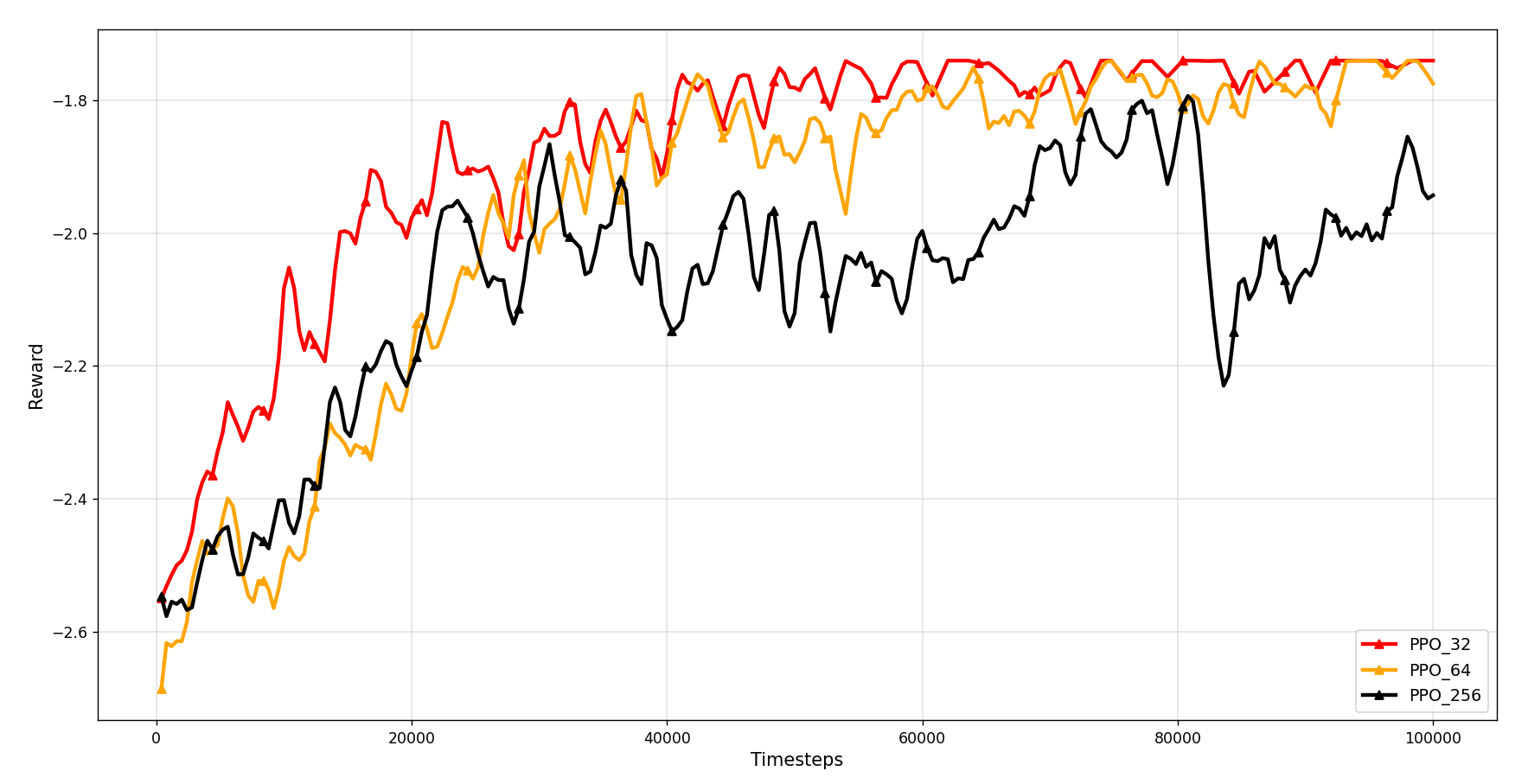}
    {Fig. 6: PPO reward for different NN architectures.}
    \label{fig:ppo-arch-reward}
\end{figure}
\begin{figure}
    \centering
    \includegraphics[width=1\linewidth]{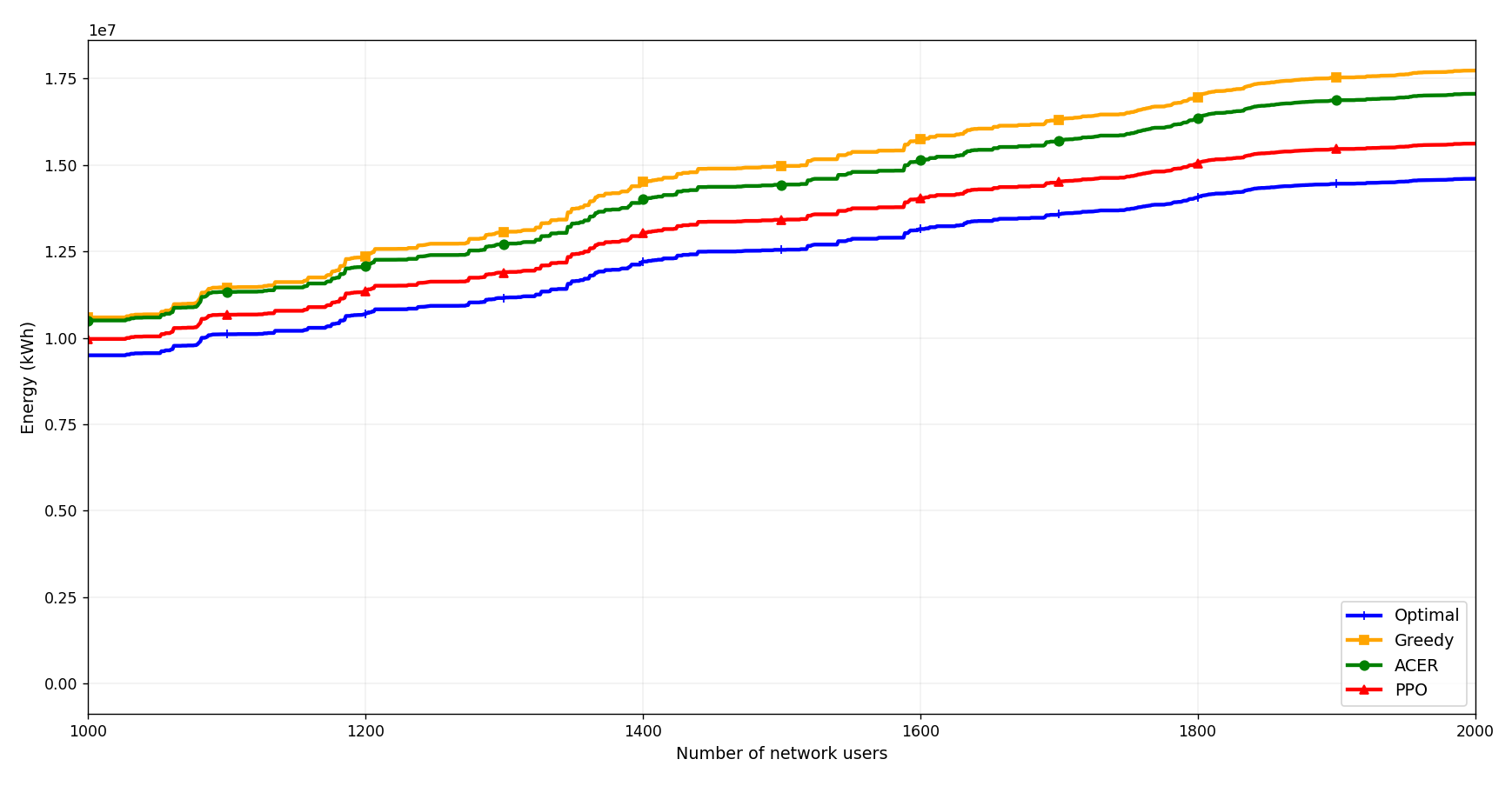}
    {Fig. 7: Energy consumption.}
    \label{fig:energy-consumption}
\end{figure}

\begin{figure}
    \centering
    \includegraphics[width=1\linewidth]{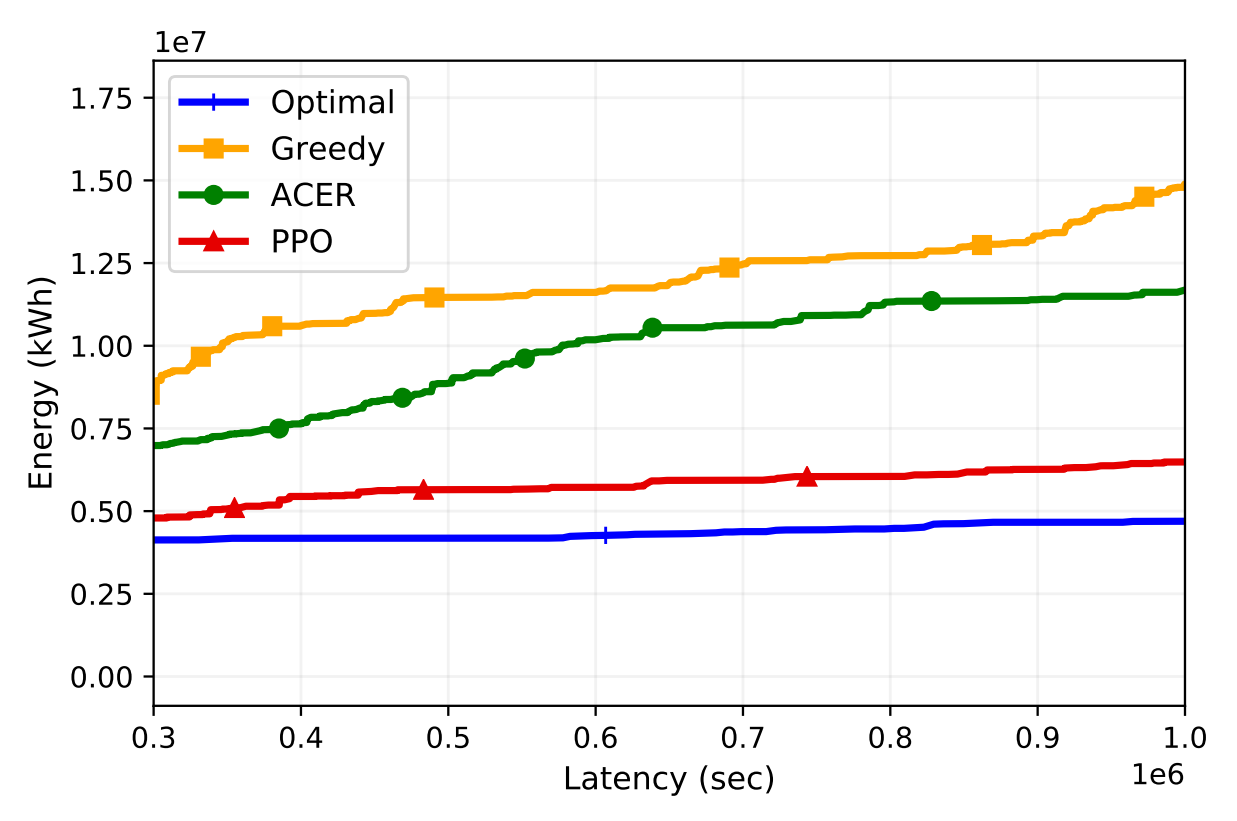}
    {Fig. 8: Trade-off between energy consumption and users latency.}
    \label{fig:energy-latency-tradeoff}
\end{figure}
 \clearpage
\section{Conclusion}
In conclusion, this paper explores the dynamics of resource allocation within Open Radio Access Networks (O-RANs) utilizing Deep Reinforcement Learning (DRL). It compares two DRL models, Proximal Policy Optimization (PPO) and Sample Efficient Actor-Critic with Experience Replay (ACER), assessing their performance in scenarios requiring real-time application and strict Quality of Service (QoS).
The replication study corroborates the original findings, demonstrating that both on-policy (PPO) and off-policy (ACER) DRL models surpass the performance of a greedy algorithm in O-RAN environments. Additionally, it confirms that PPO provides an advantageous balance between energy consumption and user latency, while ACER achieves faster convergence. These outcomes provide valuable insights into refining resource allocation strategies in O-RANs, especially in contexts like real-time surveillance video transmission where latency is crucial.
Through meticulous evaluation and comparison, the study validates the robustness and reproducibility of the proposed DRL models across various experimental setups. It also illuminates the trade-offs between energy consumption and user latency, offering a detailed understanding of the performance dynamics involved in O-RAN resource allocation.
Collectively, the findings enrich the ongoing scientific discussion, delivering practical implications for optimizing resource allocation in dynamic slicing O-RAN systems. As the O-RAN architecture continues to advance, the insights from this research are instrumental in developing more effective and intelligent network management strategies.

\end{document}